\documentclass[%
reprint,
superscriptaddress,
showpacs,preprintnumbers,
 amsmath,amssymb,
 aps,
 longbibliography,
prb,
floatfix,
]{revtex4-1}

\usepackage{xspace}
\usepackage{graphicx}
\usepackage{dcolumn}
\usepackage{bm}
\usepackage{color}
\usepackage{soul}
\usepackage{multirow}
\usepackage{hyperref}
\hypersetup{
    unicode=false,          
    pdftoolbar=true,        
    pdfmenubar=true,        
    pdffitwindow=false,     
    pdfstartview={FitH},    
    pdftitle={Phonon anomalies in FeS},    
    pdfnewwindow=true,      
    colorlinks=true,       
    linkcolor=blue,          
    citecolor=blue,        
    filecolor=blue,      
    urlcolor=blue,       
    plainpages=false,        
}

\usepackage{CJK}

\newcommand{\Ts}{\ensuremath{T_\mathrm{s}}\xspace}

\newcommand{\Tc}{\ensuremath{T_\mathrm{c}}\xspace}

\newcommand{\BFA}{\mbox{BaFe$_{2}$As$_2$}\xspace}

\newcommand{\Alg}{\texorpdfstring{\ensuremath{A_{1g}}\xspace}{A1g}}

\newcommand{\AZg}{\texorpdfstring{\ensuremath{A_{2g}}\xspace}{A2g}}
\newcommand{\Blg}{\texorpdfstring{\ensuremath{B_{1g}}\xspace}{B1g}}
\newcommand{\BZg}{\texorpdfstring{\ensuremath{B_{2g}}\xspace}{B2g}}

\newcommand{\Eg}{\texorpdfstring{\ensuremath{E_{g}}\xspace}{Eg}}
\newcommand{\grd}{$^{\circ}$\xspace}

\newcommand{\wn}{\ensuremath{\rm cm^{-1}}\xspace}
\begin{document}

\begin{CJK*}{}{}

\title{Phonon anomalies in FeS}
\date{\today}
\author{A.~Baum}
\altaffiliation{contributed equally}
\affiliation{Walther Meissner Institut, Bayerische Akademie der Wissenschaften, 85748 Garching, Germany}
\affiliation{Fakult\"at f\"ur Physik E23, Technische Universit\"at M\"unchen, 85748 Garching, Germany}
\author{A.~Milosavljevi\'{c}}
\altaffiliation{contributed equally}
\affiliation{Center for Solid State Physics and New Materials, Institute of Physics Belgrade, University of Belgrade, Pregrevica 118, 11080 Belgrade, Serbia}
\author{N.~Lazarevi\'{c}}
\affiliation{Center for Solid State Physics and New Materials, Institute of Physics Belgrade, University of Belgrade, Pregrevica 118, 11080 Belgrade, Serbia}
\author{M.~M.~Radonji\'{c}}
\affiliation{Scientific Computing Laboratory, Center for the Study of Complex Systems, Institute of Physics Belgrade, University of Belgrade, Pregrevica 118, 11080 Belgrade, Serbia}
\author{B.~Nikoli\'{c}}
\affiliation{Faculty of Physics, University of Belgrade, Studentski trg 12, Belgrade, Serbia}
\author{M.~Mitschek}
\altaffiliation{Present address: Physikalisches Institut, Goethe Universit\"{a}t, 60438 Frankfurt am Main, Germany}
\affiliation{Walther Meissner Institut, Bayerische Akademie der Wissenschaften, 85748 Garching, Germany}
\affiliation{Fakult\"at f\"ur Physik E23, Technische Universit\"at M\"unchen, 85748 Garching, Germany}
\author{Z.~Inanloo~Maranloo}
\altaffiliation{Present address: Fakult\"at f\"ur Physik E21, Technische Universit\"at M\"unchen, 85748 Garching, Germany}
\affiliation{Walther Meissner Institut, Bayerische Akademie der Wissenschaften, 85748 Garching, Germany}
\author{M.~\v{S}\'{c}epanovi\'{c}}
\affiliation{Center for Solid State Physics and New Materials, Institute of Physics Belgrade, University of Belgrade, Pregrevica 118, 11080 Belgrade, Serbia}

\author{M.~Gruji\'{c}-Broj\v{c}in}
\affiliation{Center for Solid State Physics and New Materials, Institute of Physics Belgrade, University of Belgrade, Pregrevica 118, 11080 Belgrade, Serbia}
\author{N.~Stojilovi\'{c}}
\affiliation{Center for Solid State Physics and New Materials, Institute of Physics Belgrade, University of Belgrade, Pregrevica 118, 11080 Belgrade, Serbia}
\affiliation{Department of Physics and Astronomy, University of Wisconsin Oshkosh, Oshkosh, WI 54901, USA}
\author{M.~Opel}
\affiliation{Walther Meissner Institut, Bayerische Akademie der Wissenschaften, 85748 Garching, Germany}
\author{Aifeng~Wang}
\affiliation{Condensed Matter Physics and Materials Science Department, Brookhaven National Laboratory, Upton, NY 11973-5000, USA}
\author{C.~Petrovic}
\affiliation{Condensed Matter Physics and Materials Science Department, Brookhaven National Laboratory, Upton, NY 11973-5000, USA}
\author{Z.V.~Popovi\'{c}}
\affiliation{Center for Solid State Physics and New Materials, Institute of Physics Belgrade, University of Belgrade, Pregrevica 118, 11080 Belgrade, Serbia}
\affiliation{Serbian Academy of Sciences and Arts, Knez Mihailova 35, 11000 Belgrade, Serbia}
\author{R.~Hackl}
\affiliation{Walther Meissner Institut, Bayerische Akademie der Wissenschaften, 85748 Garching, Germany}

\begin{abstract}
  We present results from light scattering experiments on tetragonal FeS with the focus placed on lattice dynamics. We identify the Raman active \Alg and \Blg phonon modes, a second order scattering process involving two acoustic phonons, and contributions from potentially defect-induced scattering. The temperature dependence between 300 and 20\,K of all observed phonon energies is governed by the lattice contraction. Below 20\,K the phonon energies increase by 0.5-1\,cm$^{-1}$ thus indicating putative short range magnetic order. Along with the experiments we performed lattice-dynamical simulations and a symmetry analysis for the phonons and potential overtones and find good agreement with the experiments. In particular, we argue that the two-phonon excitation observed in a gap between the optical branches becomes observable due to significant electron-phonon interaction.
\end{abstract}
\maketitle

\end{CJK*}


\section{Introduction}
In the iron based superconductors (IBS) magnetic order, structure, nematicity and superconductivity  are closely interrelated. Upon substituting atoms in the parent compounds the properties change in a way that the shape of the Fermi surface is generally believed to play a crucial role. Yet, the magnetic properties were found recently to be more complex and to depend also on the degree of correlation in the individual $d$ orbitals contributing to the density of states close to the Fermi surface.\cite{Yin2011_NP7_294,Si2016_NRM1_16017,Leonov2015_PRL115_106402}

The influence of correlation effects seems to increase from the 122 systems such as \BFA to the 11 chalcogenides FeTe, FeSe and FeS.\cite{Tresca2017_PRB95_205117,Miao2017_PRB95_205127} Surprisingly, the properties of the 11 class members differ substantially although they are isostructural and isoelectronic\cite{Leonov2015_PRL115_106402,Skornyakov2017_PRB96_035137}: FeSe undergoes a structural transition at \Ts $\sim 90$\,K and displays electronic nematicity.\cite{McQueen2009_PRL103_057002} While long-range magnetic order cannot be observed down to the lowest temperatures\cite{McQueen2009_PRL103_057002,Mizuguchi2010_PCS470_S338,Bendele2010_PRL104_087003,Baek2015_NM14_210} the thermodynamic properties and the Raman spectra strongly support the presence of short-ranged magnetism.\cite{He2017sep,Baum2017sep} Below \Tc $\sim 9$\,K superconductivity is observed\cite{Hsu2008_PNAS105_14262} in pristine FeSe. In mono-layer FeSe \Tc can reach values close to 100\,K.\cite{He2013_NM12_605,Ge2014_NM14_285}

The replacement of Se by Te leads to slightly off-stoichiometric Fe$_{1+y}$Te which exhibits a simultaneous magneto-structural transition near 67\,K \cite{Li2009_PRB79_054503} but is not superconducting.\cite{Fang2008_PRB78_224503,Yeh2008_EL84_37002} Finally, FeS having a superconducting transition at \Tc $\sim 5$\,K \cite{Lai2015_JACS137_10148} remains tetragonal down to the lowest temperatures.\cite{Pachmayr2016_CC52_194} It is still an open question whether tetragonal FeS hosts magnetic order. Obviously, the iron-chalcogenides are at the verge of various neighboring phases and very susceptible to small changes in the lattice and electronic structure. Yet direct access to the competing phases is still very difficult in FeTe and FeS because of the variation of the crystal quality across the families.

Here, we choose a slightly different approach and do not look directly at the electronic but rather at the lattice properties in FeS close to potential instabilities and use the Raman-active phonons as probes.
We identify the \Alg and \Blg modes, a two-phonon scattering process and a fourth mode from either defect-induced scattering or second-order scattering as well. These results are in good agreement with numerical calculations. Furthermore the temperature dependence of all phononic modes supports the results reported in Refs.~\onlinecite{Holenstein2016_PRB93_140506,Kirschner2016_PRB94_134509} where emerging short range magnetic order at approximately $20$\,K was reported.

\begin{figure}
  \centering
  \includegraphics[width=85mm]{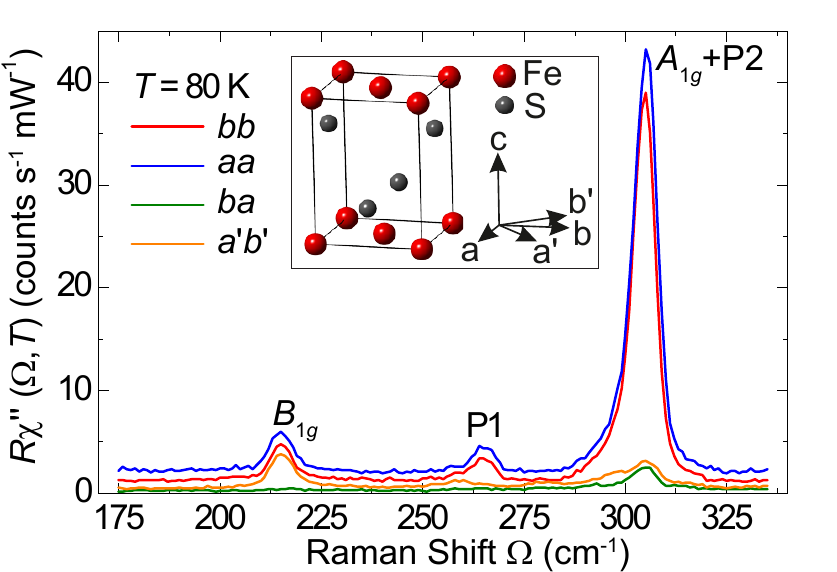}
  \caption{Raman spectra of FeS at $T = 80\,K$ measured with light polarizations as indicated. The inset shows the crystal structure of FeS and the polarization directions with respect to the crystal orientation.}
  \label{fig:assignation}
\end{figure}

\section{Experiment}
\label{sec:exp}

Single crystals of FeS were synthesized as described elsewhere.\cite{Wang2016_PRB94_094506} Before the experiment the samples were cleaved in air.

Calibrated customized Raman scattering equipment was used for the experiment. The samples were attached to the cold finger of a He-flow cryostat having a vacuum of approximately $5\cdot10^{-5}$\,Pa. For excitation we used a diode-pumped solid state laser emitting at 575\,nm (Coherent GENESIS).
Polarization and power of the incoming light were adjusted in a way that the light inside the sample had the proper polarization state and, respectively, a power of typically $P_a=3$\,mW independent of polarization. The samples were mounted as shown in the inset of Figure \ref{fig:assignation}. The crystallographic axes are $a$ and $b$ with $|a| = |b|$. The $c$-axis is parallel to the optical axis. $a^{\prime}$ and $b^{\prime}$ are rotated by 45\grd w.r.t. $a$ and $b$. The laser beam reached the sample at an angle of incidence of 66$^\circ$ and was focussed to a spot of approximately $50\,\mu{\rm m}$ diameter. The plane of incidence is the $bc$ plane. By choosing proper in-plane polarizations of the incident and scattered light the four symmetry channels \Alg, \AZg, \Blg, and \BZg of the D\textsubscript{4h} space group can be accessed. Additionally, for the large angle of incidence, exciting photons being polarized along the $b$-axis have a  finite $c$-axis projection, and the \Eg symmetry can also be accessed. For the symmetry assignment we use the 2\,Fe unit cell (crystallographic unit cell).

The observed phonon lines were analyzed quantitatively. Since the phonon lines are symmetric and $\Gamma_{\rm L}(T) \ll \omega(T)$ the intrinsic line shape can be described by a Lorentz function with a central temperature dependent energy $\omega(T)$ and a width $\Gamma_{\rm L}(T)$ (FWHM). The widths turn out to be comparable to the resolution $\sigma$ of the spectrometer. Therefore, the Lorentzian needs to be convoluted with a  Gaussian having width $\Gamma_{\mathrm{G}}\equiv\sigma$.

\section{Theory}
\label{sec:theo}

The electronic structure and the phonon dispersion were calculated using density functional theory (DFT) and  density functional perturbation theory (DFPT), respectively, \cite{Baroni2001_RMP73_515} within the QUANTUM ESPRESSO package.\cite{Giannozzi2009_JPCM21_395502} The calculations were performed with the experimental unit cell parameters $a=3.6735$\,\AA, $c=5.0328$\,\AA, and $z=0.2602$, where $z$ is the height of the sulfur atoms above the Fe plane in units of the $c$-axis.\cite{Lennie1995_MM59_677} We used the Vanderbilt ultrasoft pseudo potentials with the Becke-Lee-Yang-Parr (BLYP) exchange-correlation functional and $s$ and $p$ semi-core states included in the valence for iron. The electron-wave-function and density energy cutoffs were $70$\,Ry and $560$\,Ry, respectively, chosen to ensure stable convergence of the phonon modes. We used a Gaussian smearing of $0.01$\,Ry. The Brillouin zone was sampled with $16\times16\times16$ Monkhorst-Pack $k$-space mesh. Our electronic structure and phonon calculations are in agreement with previously reported results.\cite{Subedi2008_PRB78_134514,ElMendili2014_RA4_25827}

The experimental positions of the S atoms entail a non-zero $z$ component of the force of $6\times 10^{-2}\,{{\rm Ry}/a_{\rm B}}$ acting on them with ${a_{\rm B}}$ the Bohr radius. However, the relaxation of the $z$ positions of the S atoms would result in a large discrepancy between the calculated and experimental energies of the optical branches\cite{ElMendili2014_RA4_25827} whereas the phonon frequencies calculated from experimental structure parameters are in good agreement with the experiment (see Table~\ref{tab:LDA_phonon_energies}). When using the measured lattice parameters, including atomic positions, some of the acoustic phonons are unstable and do not have a linear dispersion at small {\bf k}. Upon relaxing the atomic positions the acoustic dispersion becomes linear, and the energies at the zone boundary decrease slightly. The energies of the optical branches, on the other hand, increase by some 10\%. Having all this in mind, we choose to use the experimental lattice parameters stated above. In this sense our calculations should be understood as a compromise.

The phonon dispersion and the density of states were calculated on a $6 \times 6 \times 6$
Monkhorst-Pack $k$-point mesh, and the dispersion is interpolated along the chosen line. The calculated phonon dispersions of the experimental and relaxed structures qualitatively coincide and display similar shapes and a gap. Discrepancies only appear in the absolute energies.

The selection rules for two-phonon processes were calculated using the modified group projector technique, (MGPT)\cite{Damnjanovic2015_PR581_1} which avoids summing over an infinite set of space group elements.

\section{Results and Discussion}
\label{sec:results}

\subsection{Polarization dependence}

Raman spectra of FeS for four linear polarization configurations at a sample temperature of $T = 80$\,K are shown in Fig.~\ref{fig:assignation}. Three peaks can be identified at 215, 265 and 305\,\wn. The symmetric peak at 215\,\wn shows up for $aa$, $bb$, and $a^\prime b^\prime$ polarizations, but vanishes for $ba$ polarization. Hence the excitation obeys \Blg selection rules and can be identified as the out-of-phase vibration of iron atoms along the $c$-axis.
The strongest slightly asymmetric peak at 305\,\wn obeys \Alg selection rules with contributions of order 5\% in $ba$ and $a^\prime b^\prime$ polarizations from either leakage or defect-induced scattering. An asymmetric Fano-type line shape can be acquired by coupling a phonon to an electronic continuum. However, as shown in Fig.~\ref{Afig:double_Voigt} in the Appendix, we find that the superposition of two symmetric, yet spectrally unresolved peaks gives a better agreement with the data than the description in terms of a Fano function. The stronger peak at 305\,\wn has \Alg symmetry with some remaining leakage. We therefore identify this mode with the in-phase vibration of sulfur atoms along the $c$-axis.
The second peak, labeled P2, appears in spectra with parallel light polarizations and vanishes in $ba$, but has some contribution in $a^\prime b^\prime$ polarizations, suggesting mixed \Alg and \Blg symmetry.
The third peak, labeled P1, is symmetric and appears only in spectra with parallel light polarizations and thus has pure \Alg symmetry.

\begin{figure}
  \centering
  \includegraphics[width=85mm]{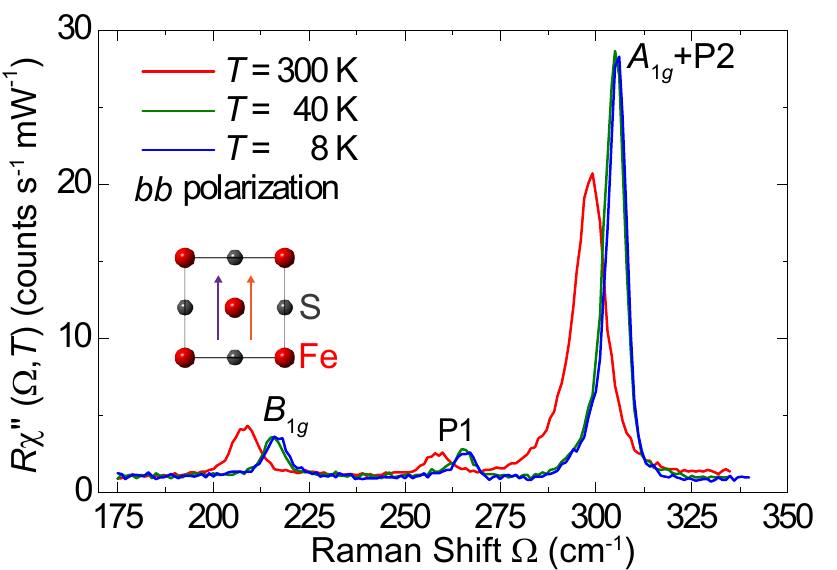}
  \caption{Raman spectra of FeS in $bb$ polarization projecting \Alg + \Blg + \Eg symmetries measured at temperatures given in the legend. The inset shows the light polarizations with respect to the crystal orientation.}
  \label{fig:data-low}
\end{figure}

\begin{figure}
  \centering
  \includegraphics[width=85mm]{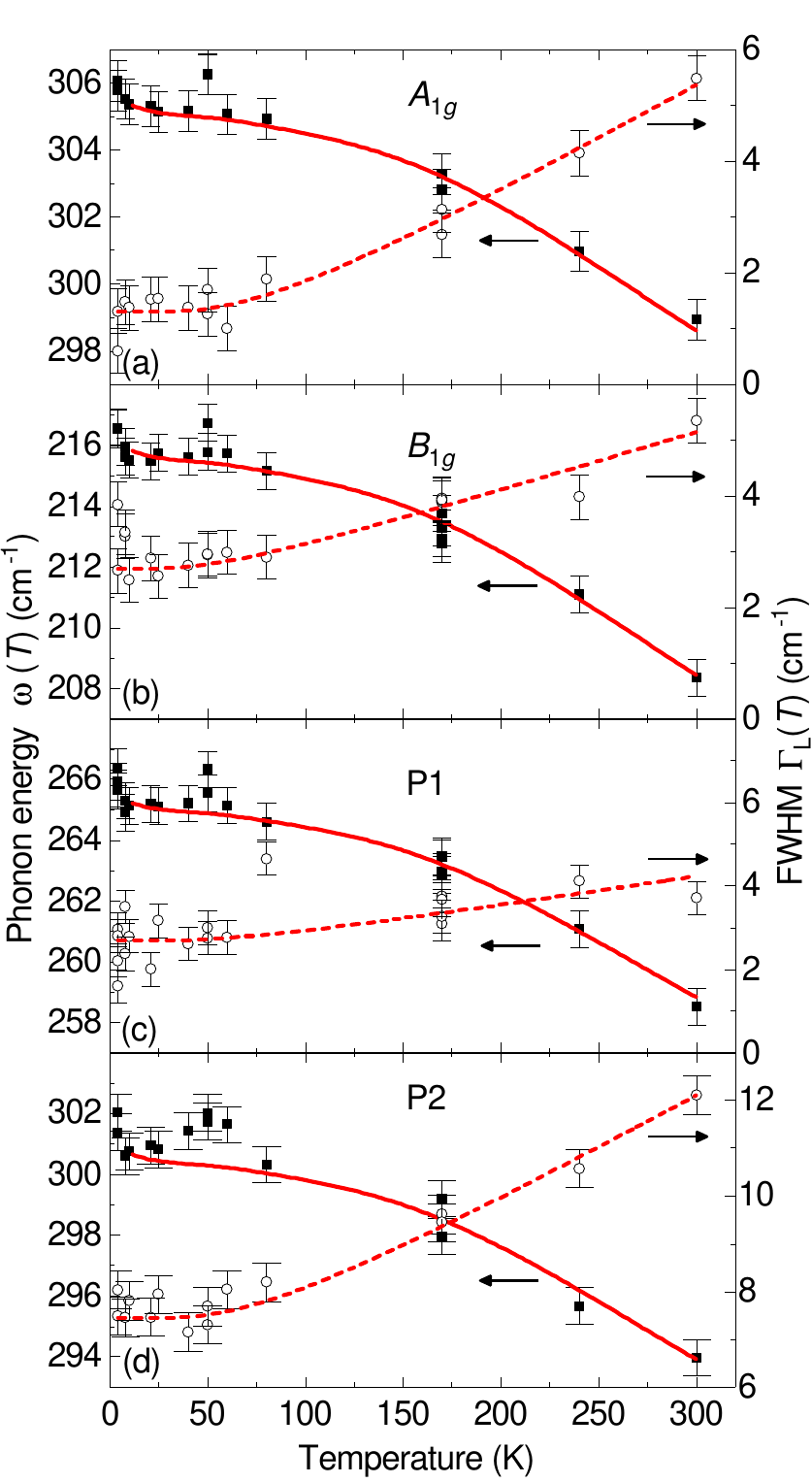}
  \caption{Temperature dependence of energy and width of the four observed phonon modes in FeS. Black squares show the phonon energies $\omega$, open circles denote the phonon line widths $\Gamma_\mathrm{L}$. The red dashed and solid lines represent the temperature dependencies of the phonon line widths and energies according to Eqs.~\eqref{eq:FWHM_fitfunc} and \eqref{eq:energy_fitfunc}, respectively. For better visualizing the low-temperature part, the data of this figure are plotted on a logarithmic temperature scale in Fig.~\ref{Afig:temperaturedep_log} of Appendix~\ref{Asec:temperaturedep_log}.}
  \label{fig:temperaturedep}
\end{figure}

\subsection{Temperature dependence}
\label{sec:T}

For properly assigning all observed modes and for getting access to putative phase transitions we studied the temperature dependence. Figure~\ref{fig:data-low} shows Raman spectra in $bb$-polarization at 8, 40, and 300\,K. The three peaks shift to higher energies upon cooling. The fourth peak P2 can not be resolved in the raw data and can only be analyzed after a fitting procedure (see Appendix ~\ref{Asec:double_Voigt_fit}). The peak energies $\omega(T)$ and the (intrinsic) line widths $\Gamma_\mathrm{L}(T)$ were determined as described at the end of section \ref{sec:exp}. All four modes show a monotonous increase in energy and decrease in line width upon cooling as shown in Fig.~\ref{fig:temperaturedep}. Below  20\,K the increase in the energies accelerates. We first address this overall behavior and disregard the anomaly around 50\,K for the moment.

The shift and narrowing of all modes can be explained in terms of lattice contraction using a constant Gr\"uneisen parameter $\gamma$ and anharmonic decay into other phonon modes, respectively. The change in the (Lorentzian) line width  $\Gamma_\mathrm{L}(T)$ is given by,\cite{Klemens1966_PR148_845}
\begin{equation}
  \Gamma_\mathrm{L}(T) = \Gamma_{\mathrm{L},0} \left(1 + \frac{2    \lambda_{\rm{ph-ph}}}{\exp\left(\frac{\hbar \omega_0}{2 k_{\rm B} T}\right)-1}\right).
\label{eq:FWHM_fitfunc}
\end{equation}

The zero temperature limits $\Gamma_{\mathrm{L},0}$ and $\omega_0$ were obtained by extrapolating the respective experimental points of $\Gamma_{\mathrm{L}}(T)$ and $\omega(T)$ in the range $20\le T \le 50$\,K to $T=0$  (Fig.~\ref{fig:temperaturedep}). With the phonon-phonon coupling $\lambda_{ph-ph}$ being the only free parameter the temperature dependence of $\Gamma_{\mathrm{L}}(T)$ can be described as shown by red dashed lines in Fig.~\ref{fig:temperaturedep}. The phonon energy $\omega(T)$ contains contributions from both the anharmonic decay and the lattice contraction, which depends essentially on the thermal occupation of the phonons, and can be written as\cite{Eiter2014_PRB90_024411}
\begin{eqnarray}
  \omega(T) = \omega_0\left[\frac{}{}\right.\!\!1
                             &-& \gamma  \frac{V(T)-V_0}{V_0} \nonumber \\
                             &-&\!\left.\left(\!\frac{\Gamma_{\mathrm{L},0}}
                             {\sqrt{2}\omega_0}\!\right)^{\!2}\!\! \left(1 \!+\!
                             \frac{4 \lambda_{\rm{ph-ph}}}{\exp\left(\frac{\hbar \omega_0}
                             {2 k_{\rm B} T}\right) \!-\!1}\right)\right]\!.
\label{eq:energy_fitfunc}
\end{eqnarray}

$V(T)$ and $V_0$ are the volumes of the unit cell at temperatures $T$ and $T\to0$, respectively. The numbers for the calculations are taken from Ref.~\onlinecite{Pachmayr2016_CC52_194}. The second term describes the effect of phonon damping on the line position in the harmonic approximation. Using $\lambda_{\rm ph-ph}$ from Eq.~\ref{eq:FWHM_fitfunc},  the Gr\"{u}neisen parameter $\gamma$ is the only free parameter and is assumed to be constant. The temperature dependence $\omega(T)$ resulting from the fits are plotted in Fig.~\ref{fig:temperaturedep} as solid red lines. The numerical values for parameters $\gamma$ and $\lambda_{\rm ph-ph}$ obtained from the $T$-dependent energy and line width are compiled in Table~\ref{tab:fit_parameters}.

\begin{table}
  \caption{Symmetry, Gr\"{u}neisen constant $\gamma$, and phonon-phonon coupling parameter $\lambda_{\rm{ph-ph}}$ of the four experimentally observed modes.}
	\label{tab:fit_parameters}
  \renewcommand{\arraystretch}{1.5}
	\begin{ruledtabular}
	\begin{tabular}{cccc}
		Mode & Symmetry   & $\gamma$ & $\lambda_{\rm{ph-ph}}$ \\ \hline
		S    & \Alg	      &	2.2      & 1.68 \\
		Fe   & \Blg 	  & 3.4      & 0.31 \\
		P1   & \Alg       & 2.4	     & 0.25 \\
		P2   & \Alg+ \Blg & 2.2      & 0.31 \\
	\end{tabular}
	\end{ruledtabular}
\end{table}

\begin{table}
  \caption{Raman active phonon modes in t-FeS. Shown are the symmetries, the theoretical predictions for the experimental lattice parameters at $T=0$, and the atoms involved in the respective vibrations. The experimental energies in the third column are extrapolations to $T = 0$ of the points measured between 20\,K and 50\,K.}
	\label{tab:LDA_phonon_energies}
  \renewcommand{\arraystretch}{1.5}
	\begin{ruledtabular}
	\begin{tabular}{cccc}
		\multirow{2}{*}{Symmetry} & \multicolumn{2}{c}{Phonon energy (\wn)} &
        \multirow{2}{2cm}{\centering Atomic displacement} \\
		              & Calculation & Experiment & \\  \hline
		\Alg	      &	316.1       & 305.3      & S \\
		\Blg	      & 220.4	    & 215.8      & Fe \\
		\Eg		      &	231.6	    &    	     & Fe, S \\
		\Eg		      &	324.8	    &		     & Fe, S \\
	\end{tabular}
	\end{ruledtabular}
\end{table}

Below 20\,K and around 50\,K anomalies are found in the experimental data:\\
(\textit{i}) At 50\,K the peak energies of all four modes deviate significantly from the otherwise smooth temperature dependence. The nearly discontinuous increase in energy could be reproduced for the \Alg phonon and peak P2 in multiple measurements. For the \Blg phonon and  mode P1 the anomaly is not as clearly reproducible. The energy anomalies do not have a correspondence in the line width. As there is neither an abrupt change in the lattice constants \cite{Pachmayr2016_CC52_194} nor any other known phase transition close to 50\,K the origin of this anomaly remains unexplained although we consider it significant.\\
(\textit{ii}) Upon cooling from 20\,K to 4\,K all four modes exhibit sudden, yet small, increases in energy. The changes in width are heterogeneous in that  the \Alg mode narrows and the \Blg mode broadens. No clear tendencies can be derived for modes P1 and P2. Sudden changes in the temperature dependence typically indicate phase transitions. Yet, no phase transition has been identified so far. However, the anomaly at 20\,K coincides with the emergence of short range magnetic order as inferred from two $\mu \rm SR$ studies.\cite{Holenstein2016_PRB93_140506,Kirschner2016_PRB94_134509} Susceptibility measurements on a sample from the same batch were inconclusive. On the other hand, the XRD data show a small anomaly in the lattice parameters, and the unit cell volume does not saturate at low temperature but rather decreases faster between 20\,K and 10\,K than above 20\,K.\cite{Pachmayr2016_CC52_194} This volume contraction by and large reproduces the change in the phonon energies as can be seen by closely inspecting the low-temperature parts of Fig.~\ref{fig:temperaturedep} (see also Fig.~\ref{Afig:temperaturedep_log}). Hence, the indications of short-range magnetism in FeS found by $\mu \rm SR$ have a correspondence in the temperature dependence of the volume and  the phonon energies.

Clear phonon anomalies were observed at the onset of the spin density wave (SDW) phases in 122 systems\cite{Rahlenbeck2009_PRB80_064509,Chauviere2009_PRB80_094504,Chauviere2011_PRB84_104508} and of the more localized magnetic phase in FeTe\cite{Um2012_PRB85_064519} whereas continuous temperature dependence of the phonons was found in systems without long-range magnetism.\cite{Um2012_PRB85_012501,Gnezdilov2013_PRB87_144508}
Upon entering the SDW state in the 122 systems the \Alg (As) mode softens abruptly and narrows by a factor of three whereas the \Blg (Fe) mode stays pinned and narrows only slightly.\cite{Rahlenbeck2009_PRB80_064509} The strong coupling of the As mode to magnetism was traced back to the interaction of the Fe magnetic moment with the Fe-As tetrahedra angle\cite{Yildirim2009_PRL102_037003} which goes along with a change of the $c$-axis parameter. In Fe$_{1+y}$Te the roles of the \Blg and \Alg modes are interchanged.\cite{Gnezdilov2011_PRB83_245127,Um2012_PRB85_064519,Popovic2014_SSC193_51} In contrast, all four modes observed here in  FeS harden below $T^\ast \approx 20\,K$ being indicative of a type of magnetic ordering apparently different from that in the other Fe-based systems.

Very recently commensurate magnetic order with a wave vector of \textbf{q} = (0.25,0.25,0) was found in FeS below $T_{\rm N} = 116$\,K using neutron powder diffraction.\cite{Kuhn2017_PC534_29} In the Raman spectra no anomalies can be seen around 120\,K even if the range is studied with fine temperature increments of 10\,K as shown in Appendix \ref{Asec:high_temperature}. However, a small change in the temperature dependence of the $c$-axis parameter is observed around 100\,K by XRD\cite{Pachmayr2016_CC52_194} which could be related to this type of magnetic order. Since the influence on the volume is small there is no detectable impact on the phonons.

\subsection{Analysis of the modes P1 and P2}

Based on the energies, the selection rules, and the temperature dependence we first clarify the phononic nature of the two lines P1 and P2 which cannot as straightforwardly be identified as lattice vibrations as the in-phase sulfur and out-of-phase iron vibrations at 305.3 and 215.8\,\wn. Second we derive their origin from the phonon density of states (PDOS) calculated for the zero-temperature limit.

All experimental energies for $T \to 0$ were derived from the points at low temperature as described in subsection~\ref{sec:T}  (see also Fig.~\ref{fig:temperaturedep}). The results for the modes at the $\Gamma$ point are summarized in Table~\ref{tab:LDA_phonon_energies} and can be directly compared to the results of the calculations. The discrepancies between the experimental and theoretical energies for the Raman-active phonons are smaller than 4\%. The price for this accuracy in the optical energies is an instability and possibly too high energies in the acoustical branches at small and, respectively, large momentum (see section~\ref{sec:theo}).

The unidentified peaks P1 and P2 appear in the spectra measured with $aa$ polarization, where none of the electric fields has a projection on the $c$-axis. Thus they cannot have \Eg symmetry obeying $ca$ and $cb$ selection rules. In addition, the observed energies would be relatively far off of the calculated energies (see Table~\ref{tab:LDA_phonon_energies}).
Both peaks exhibit  temperature dependencies similar to those of the two Raman-active phonons, and the Gr\"{u}neisen parameters are close to the typical value\cite{MacDonald1981_PRB24_1715} of 2 and similar to those of the Raman-active phonons. The phonon-phonon coupling parameters $\lambda_{\rm{ph-ph}}$ derived from the temperature dependence of the line widths are close to 0.3 similar to that of the \Blg phonon. $\lambda_{\rm{ph-ph}}$ of the \Alg phonon is roughly six times bigger for reasons we address later. Yet, because of the small prefactor $\left({\Gamma_{\mathrm{L},0}}/\sqrt{2}{\omega_0}\right)^2 = {\cal O}(10^{-3})$ the contribution of phonon-phonon coupling to the temperature dependence of $\omega(T)$ remains negligible in all cases, and the phonon energies are essentially governed by the lattice contraction. These considerations demonstrate the phononic origin of the peaks P1 and P2.

\begin{figure}
  \centering
  \includegraphics[width=85mm]{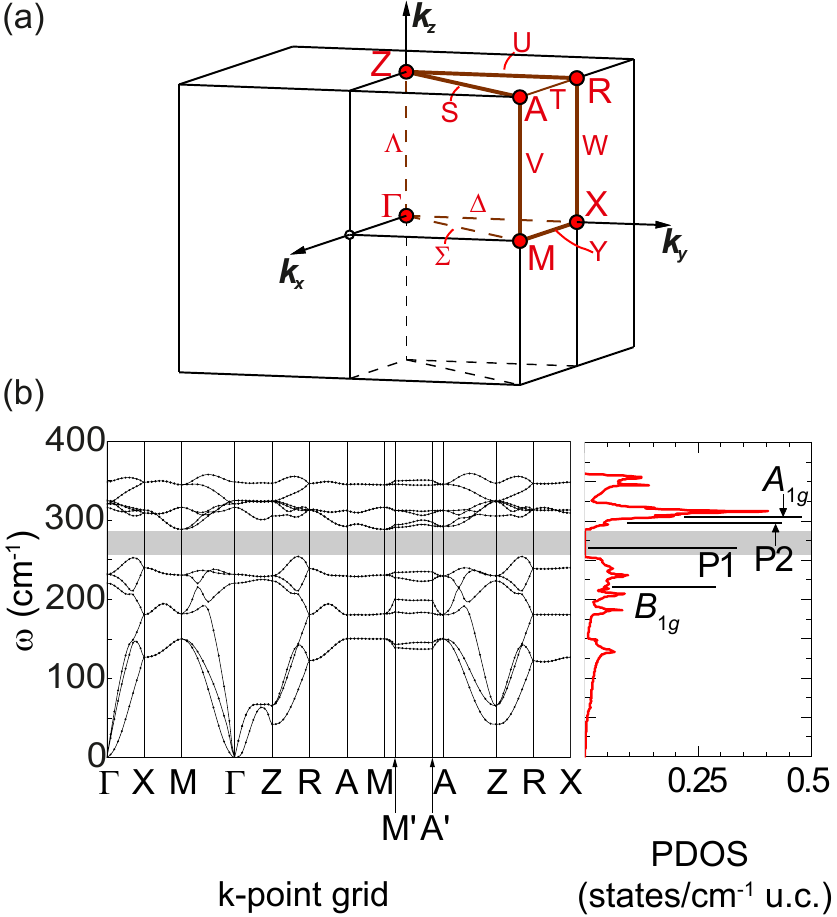}
  \caption{Phonon dispersion of t-FeS. (a) Brillouin zone with high symmetry points and lines.\cite{Aroyo2014_ACSA70_126} (b) Phonon dispersion along the directions as indicated and phonon density of states (PDOS). The gray-shaded area marks the gap in the phonon dispersion. The dispersion shown here is derived using experimental lattice parameters. For this reason some of the acoustic phonons are unstable and do not have a linear dispersion around the $\Gamma$ point. Upon relaxing the structure the acoustic dispersion becomes linear at $\Gamma$, and the energies at the zone boundary decrease slightly. The energies of the optical branches, on the other hand, increase by some 10\%. M'=(0.4, 0.4, 0.0) and A'=(0.4, 0.4, 0.5). The experimental energies of the four observed modes are shown as black lines.
  }
  \label{fig:BZ_dispersion}
\end{figure}

In the second step we try to identify the phonon branches which P1 and P2 can be related to. To this end the full phonon dispersion and density of states (PDOS) were derived as described in section \ref{sec:theo} and are plotted in Fig. \ref{fig:BZ_dispersion}.

Independent of using the relaxed or experimental structure, P1 is located in the gap of the (theoretical) PDOS and cannot result from first order defect-induced Raman scattering. What alternatives exist for explaining P1? If we exclude exotic explanations such as a collective mode for the reasons given above the energy of $\omega_{\rm P1}= 265$\,\wn can only be obtained by the sum of two phonon modes having equal energy $\omega_{\rm P1}/2$ and momenta {\bf k} and {-\bf k} (for maintaining the $q\approx0$ selection rule). As shown for various transition metal compounds including TiN, ZrN or NbC second-order phonon Raman scattering can occur in the presence of defects.\cite{Spengler1976_SSC18_881} Then first-order scattering being proportional to the PDOS (modulo energy and symmetry dependent weighting factors) is expected to be also substantial if not stronger. Although our crystals are slightly disordered there is no indication of substantial intensity at energies with high PDOS as can be seen by directly comparing Figs.~\ref{fig:assignation} and \ref{fig:BZ_dispersion}\,(b). Alternatively, second-order scattering can originate in enhanced electron-phonon coupling.\cite{Spengler1978_PRB17_1095} In either case the energies of two phonons add up as they get excited in a single scattering process.
Generally, no selection rules apply for second order Raman scattering and the resulting peak would appear in all symmetry channels.\cite{Hayes1978_Scattering} Exceptions exist if the phonon wave-vectors coincide with high-symmetry points or lines of the Brillouin zone.

From the phonon dispersion alone several phonon branches having {\bf k} and {-\bf k} and energies in the range around $\omega_{\rm P1}/2$ could add up to yield 265\,\wn [see Fig.~\ref{fig:BZ_dispersion}]. However, as explained in Appendix~\ref{Asec:MGPT} and shown in Table~\ref{table:MGPT} for the space group P4/nmm of t-FeS, the \Alg selection rules of P1 exclude all non-symmetric combinations of branches (right column of Table~\ref{table:MGPT}). On the other hand, all symmetric combinations include \Alg selection rules for the two-phonon peak (left column of Table~\ref{table:MGPT}), and one has to look essentially for a high PDOS in the range $\omega_{\rm P1}/2$. As shown in Fig.~\ref{fig:BZ_dispersion}\,(b) the PDOS has a maximum in the right energy range. Since the maximum results from momenta away from the high-symmetry points or lines (see Fig.~\ref{fig:BZ_dispersion}) which alone lead to pure \Alg symmetry one expects also intensity in \Blg and \Eg symmetry as opposed to the experiment. For exclusive \Alg selection rules only seven possibilities exist. Since phase space arguments favor modes having a flat dispersion in extended regions of the Brillouin zone the $\Gamma$, $M$ and/or $A$ points are unlikely to give rise to P1, and only the lines $S= A-Z$, $\Sigma=\Gamma-M$, and $V=A-M$ remain. The dispersion along the $S$ or $\Sigma$ branch contributes very little to the PDOS. On the high-symmetry line $V$ a doubly degenerate branch would have a flat dispersion [see Fig.~\ref{fig:BZ_dispersion}\,(b)] and contribute substantially to the PDOS but the energy of 150\,\wn differs by 13\% from the expected energy of 132.5\,\wn. Instead of arguing about the accuracy of the theoretical phonon energies (see section~\ref{sec:theo}) we looked at the dispersion close to but not strictly on $V$ where the contribution to \Blg and \Eg symmetries is expected to be still very small, e.g. along $M^\prime-A^\prime$ [Fig.~\ref{fig:BZ_dispersion}(b)]. A detailed inspection shows that the maximum of the PDOS between 130 and 140\,\wn comes from there. This explains both the selection rules and the energy of P1 to within a few percent.

Peak P2 cannot be explained in terms of one of the two \Eg phonons either. As opposed to P1 it is not inside the gap of the PDOS and thus can originate from either first or second order scattering. If P2 originates in second order scattering in the same fashion as P1 there are five possibilities yielding \Alg+\Blg but not \Eg selection rules. As explained in the last paragraph only the branches $\Delta = \Gamma - X$ and $U = Z -R$ may contribute.
For the low PDOS there we consider also first order defect-induced scattering for P2 to originate from. In fact, the PDOS possesses its strongest maximum 5\,\wn below the (theoretical) \Alg phonon exactly where P2 is found. In spite of the very high PDOS here, the peak is weak explaining the negligible contributions from first order defect-induced scattering at lower energies. The high PDOS between 300 and 325\,cm$^{-1}$ may also be an alternative yet less likely explanation for the weak contributions in crossed polarizations in the energy range of the \Alg phonon (Fig.~\ref{fig:assignation}).

Finally, we wish to clarify whether the large phonon-phonon coupling $\lambda_{\rm ph-ph}^{A1g}$ found for the \Alg Raman-active mode (see Table~\ref{tab:fit_parameters}) is related to the appearance of P1. Due to the close proximity of the energies the \Alg mode apparently decays into states close to those adding up to yield  P1. The decay is less restricted by symmetry leaving more options. For both processes the phonon-phonon coupling  has to be substantial with the order of magnitude given by $\lambda_{\rm ph-ph}^{A1g}\approx 1.7$. Phonon-phonon coupling is present in any type of material because of the anharmonic potential. Defects enhance this effect \cite{Spengler1976_SSC18_881}. Since FeS is a metal the phonon-phonon coupling goes at least partially through electronic states and may be indicative of enhanced electron-phonon coupling, $\lambda_\mathrm{el-ph}$, as described, e.g., in Ref.~\onlinecite{Spengler1978_PRB17_1095}. The related contribution to $\lambda_\mathrm{ph-ph}$ is then expected to be proportional to $\lambda_\mathrm{el-ph}^2$. This conclusion is compatible with early results on the branch-dependent electron-phonon coupling in \mbox{LaFeAsOF} where the strongest effects are reported for some $\Gamma$-point modes and the acoustic branches with intermediate to large momenta.\cite{Boeri2008_PRL101_026403} $\lambda_{\rm ph-ph}^{A1g} > 1$ and the two-phonon peak P1 indicate that the electron-phonon coupling is possibly larger than in the other Fe-based systems and reaches values up to unity. In \BFA, as an example from the pnictide family, $\lambda_{\rm el-ph}^2\approx 1\dots4\times10^{-2}<\lambda_{\rm ph-ph}\approx 0.1$ is reported \cite{Mansart2010_PRB82_024513,Rettig2013_NJP15_083023,Rahlenbeck2009_PRB80_064509}. On the other hand, one finds $\lambda_{\rm el-ph}^2\approx 0.4<\lambda_{\rm ph-ph}\approx 0.9$ for the $E_g$ phonon in MgB$_2$, being generally believed to be a conventional superconductor \cite{Martinho2003_SSC125_499,Wang2001_PCS355_179}. Thus, one may speculate whether $\lambda_{\rm el-ph}$ might be even large enough in FeS to account for a \Tc in the 5\,K range.

\section{Conclusion}
\label{sec:conclusion}

We have studied and identified phonons in tetragonal FeS by Raman scattering. For the \Alg sulfur and \Blg iron mode the DFT and  DFPT calculations agree to within  a few percent with the experiment. A third observed peak within a gap in the theoretical phonon density of states can be identified as a second order scattering process involving two phonons. Both the selection rules, based on the modified group projector technique, and the energy are in agreement with experiment. A fourth mode identified close to the \Alg sulfur can be traced back to the biggest maximum of the PDOS and is most likely activated by a small amount of defects.

The temperature dependence of all four modes is governed by the contraction of the lattice, but shows anomalies at 50\,K and below 20\,K. The anomaly observed at 20\,K has a correspondence in the thermal expansion \cite{Pachmayr2016_CC52_194} and $\mu$SR experiments\cite{Holenstein2016_PRB93_140506, Kirschner2016_PRB94_134509} which indicate short-range magnetic order. The long-range magnetic order observed recently by neutron diffraction experiments\cite{Kuhn2017_PC534_29} below $T_{\rm N}$=116\,K has no correspondence in the Raman spectra.

The appearance of two-phonon scattering indicates strong phonon-phonon scattering which is likely to originate from an electron-phonon interaction being enhanced in comparison to other pnictides and chalcogenides. We argue that in FeS the \Tc can in principle entirely result from electron-phonon interaction.

\section*{Acknowledgement}
We acknowledge valuable discussions with T. B\"ohm and D. Jost. The work was supported by the German Research Foundation (DFG) via the Priority Program SPP\,1458 (grant-no. Ha2071/7) and the Serbian Ministry of Education, Science and Technological Development under Projects III45018 and ON171017. Numerical simulations were run on the PARADOX supercomputing facility at the Scientific Computing Laboratory of the Institute of Physics Belgrade. We acknowledge support by the DAAD through the bilateral project between Serbia and Germany (grant numbers 56267076 and 57142964). Work carried out at the Brookhaven National Laboratory was primarily supported by the Center for Emergent Superconductivity, an Energy Frontier Research Center funded by the U.S. DOE,
Office of Basic Energy Sciences (A.W. and C.P.). N.S. was supported by UW Oshkosh FDS498 grant.



%



\clearpage
\begin{appendix}
\label{sec:appendix}

\setcounter{figure}{0}
\renewcommand\thefigure{A\arabic{figure}}

\setcounter{table}{0}
\renewcommand\thetable{A\Roman{table}}

\section{Magnetization measurements}
\label{Asec:magnetization}
\begin{figure}
  \centering
  \includegraphics[width=85mm]{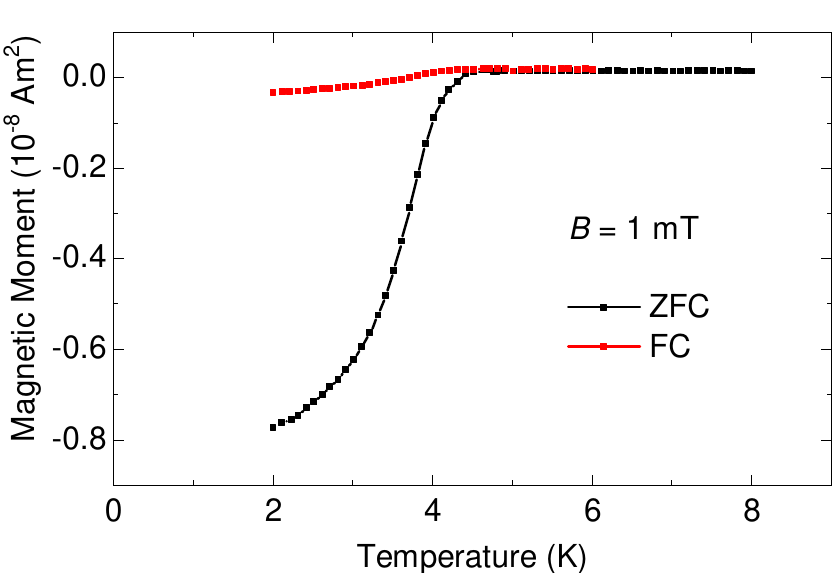}
  \caption[]{(Color online) Magnetization measurements of \mbox{t-FeS} at an applied field of $B = 1\,\mathrm{mT}$ cooled to 2\,K with (red curve) and without applied field (black curve)}
  \label{Afig:magnetization}
\end{figure}

Fig.~\ref{Afig:magnetization} shows magnetization measurements on a \mbox{t-FeS} sample from the batch studied in small applied fields. Measurements were done on a Quantum Design MPMS XL-7 SQUID magnetometer by cooling the sample to 2\,K and sweeping the temperature at 0.1\,K/min. When cooled without applied field (ZFC, black curve) the sample shows a superconducting transition with onset at 4.5\,K and a center of the transition at 3.6\,K. When cooled in an applied field the magnetization decreases only weakly in the superconducting state indicating strong pinning.

\section{Decomposition of the line at \texorpdfstring{305\,\wn}{305 1/cm}}
\label{Asec:double_Voigt_fit}
\begin{figure}
  \centering
  \includegraphics[width=85mm]{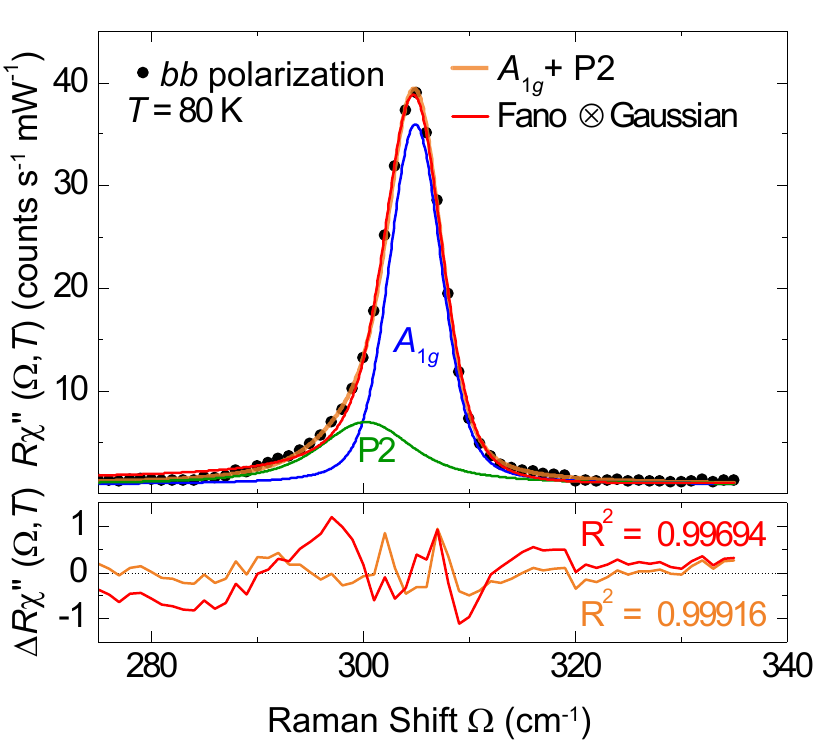}
  \caption[]{(Color online) Decomposition of the asymmetric phonon peak at 305\,\wn. Measured data are shown as black dots. The orange line shows the sum of two Voigt profiles shown as blue and green lines, respectively. The convolution of Fano and Gaussian (red line) deviates in the peak flanks and the nearby continuum.}
	\label{Afig:double_Voigt}
\end{figure}

The peak at 305\,\wn at low temperatures shows a significant asymmetry towards lower energies (see also Fig.~\ref{fig:assignation}). Coupling of the \Alg phonon mode to an electronic continuum by strong electron-phonon coupling would result in a line shape given by the convolution of a Fano function and a Gaussian, the latter representing the resolution of the spectrometer. We find, however, that this does not yield a satisfactory description of the measured line shape as can be seen from the red curve in Fig. \ref{Afig:double_Voigt}, and thus conclude that the asymmetry of the peak stems from the overlap of two peaks which cannot be resolved separately. The corresponding line shape is the sum of two Lorentzians each convoluted with a Gaussian which governs the resolution of the setup. Due to the distributivity of the convolution this is identical to the sum of two Voigt functions sharing the same width $\Gamma_\mathrm{G}$ of the Gaussian part. The overall spectral shape is shown in Fig.~\ref{Afig:double_Voigt} as orange line and agrees excellently with the data. The two contributing lines are shown in blue and green. From the selection rules (see Fig.~\ref{fig:assignation}) we identify the blue curve as the in-phase vibration of sulfur atoms in \Alg symmetry. The green line denotes a second mode P2, the origin of which is discussed in the main text.

\section{Detailed temperature dependence for \texorpdfstring{$80 \le T \le 300$\,K}{80 < T < 300 K}}
\label{Asec:high_temperature}
\begin{figure}
  \centering
  \includegraphics[width=85mm]{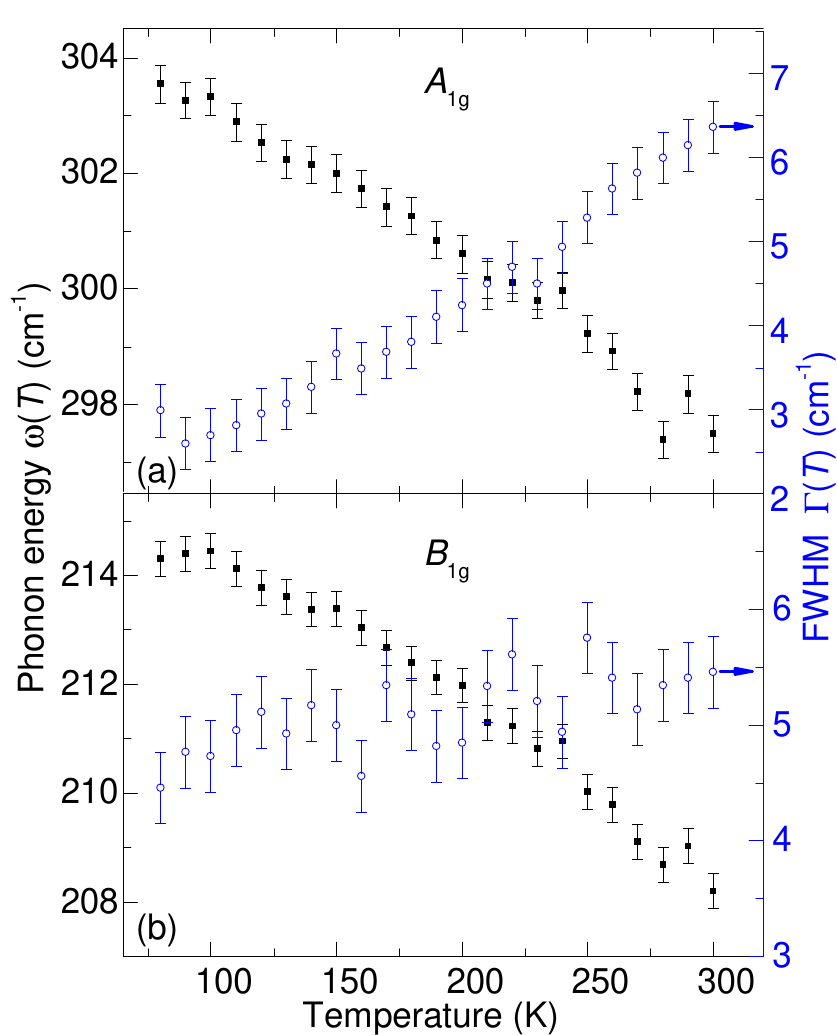}
  \caption[]{(Color online) Temperature dependence of \Alg and \Blg phonon modes in the temperature range between 80\,K and 300\,K. Black squares denote the phonon energies, open circles denote the phonon line widths.}
  \label{Afig:no_effect}
\end{figure}

Fig.~\ref{Afig:no_effect} shows the temperature dependence of the energies $\omega$ and line widths $\Gamma(T)$ (FWHM) from 80\,K to 300\,K measured in temperature increments of 10\,K.
Raman scattering measurements were performed using a Jobin Yvon T64000 Raman system in micro-Raman configuration. A solid state laser with 532 nm line was used as an excitation source. Measurements were performed in high vacuum ($10^{-6}$\, mbar) using a KONTI CryoVac continuous Helium flow cryostat with 0.5 mm thick window.  Laser beam focusing was accomplished using a microscope objective with $\times$50 magnification. The samples were cleaved right before being placed in the vacuum. As can be seen from Fig.~\ref{Afig:no_effect}, there is no deviation from the standard temperature behavior around 120\,K.

\section{Temperature dependence on a logarithmic scale}
\label{Asec:temperaturedep_log}
\begin{figure}
  \centering
  \includegraphics[width=85mm]{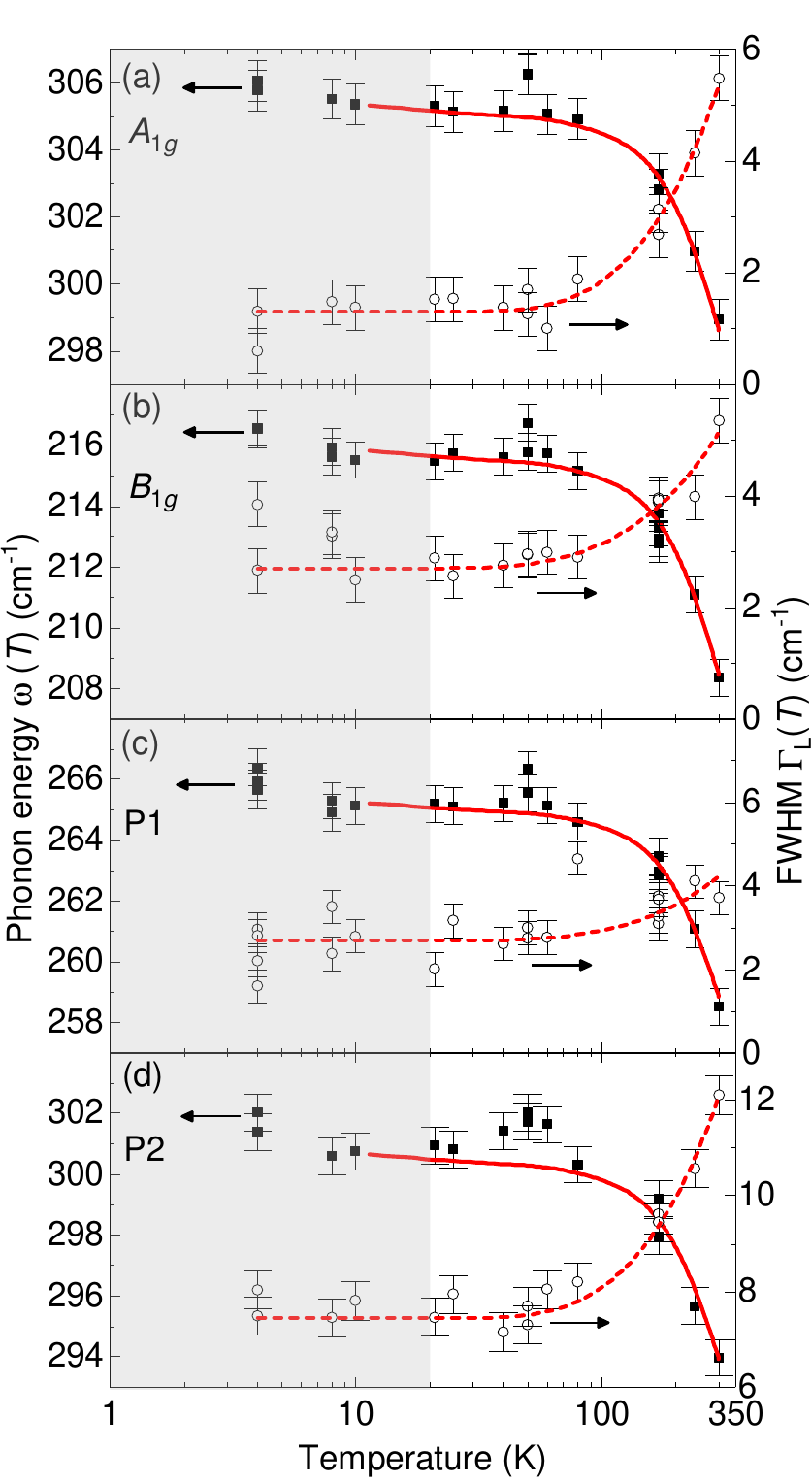}
  \caption[] (Color online) Temperature dependence of energy and width of the four observed phonon modes in FeS on a logarithmic scale. The data is identical to Fig.~\ref{fig:temperaturedep} of the main text. Black squares show the phonon energies $\omega$, open circles denote the phonon line widths $\Gamma_\mathrm{L}$. The red dashed and solid lines represent the temperature dependence of the phonon line widths and energies according to Eqs.~\eqref{eq:FWHM_fitfunc} and \eqref{eq:energy_fitfunc}, respectively. The region below 20\,K is shaded light gray. Since the data for the volume are limited to the range above 10\,K the theoretical curves for the phonon energies (full red lines) end at 10\,K
	\label{Afig:temperaturedep_log}
\end{figure}

To better illustrate the behavior of the phonons at low temperatures Fig.~\ref{Afig:temperaturedep_log} shows the experimental data and the theoretical curves from Fig.~\ref{fig:temperaturedep} of the main text on a logarithmic temperature scale. The region below 20\,K is shaded light gray. As explained in section~\ref{sec:T} all four modes show an increase in energy below 20\,K instead of the expected saturation, indicative of the putative onset of short range magnetic order. This effect manifests itself also in an incipient decrease of the unit cell volume \cite{Pachmayr2016_CC52_194} and is visible in the theoretical results for the phonon energies (full red lines). No clear tendency can be seen for the line widths. The energy anomaly found around 50\,K is discussed in the same section.

\section{Second sample batch}
\label{Asec:second_sample}
\begin{figure}
  \centering
  \includegraphics[width=85mm]{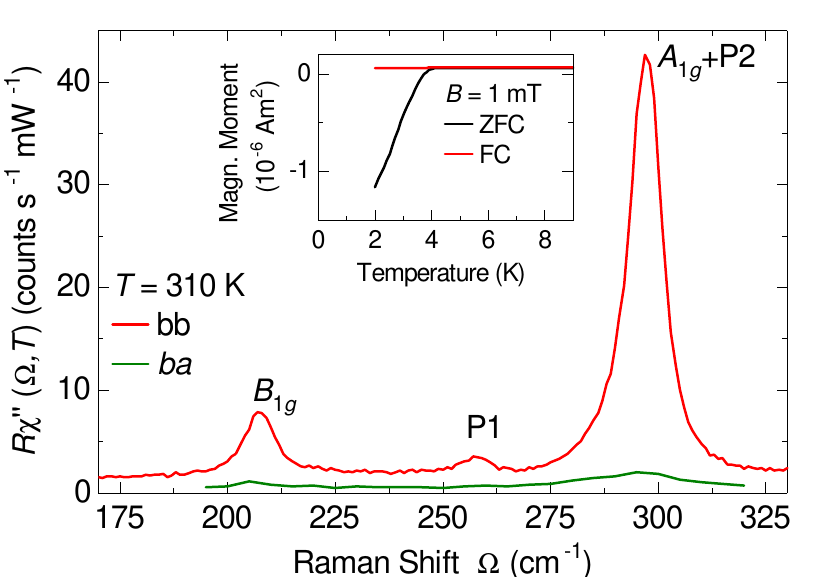}
  \caption[]{(Color online) Raman spectra of a t-FeS sample from a different batch taken at $T$ =310\,K in polarizations as given in the legend. The inset shows magnetization measurements on a sample from this batch similar to Appendix \ref{Asec:magnetization}}.
	\label{Afig:second_sample}
\end{figure}

Fig.~\ref{Afig:second_sample} shows Raman spectra on a t-FeS sample from a different batch (E256) taken at $T$ =310\,K. The sample was oriented the same way as described in the main text. All three modes are visible for parallel light polarizations ($bb$), but vanish for crossed polarizations ($ba$), confirming the selection rules observed in the sample described in the main text. The inset shows magnetization measurements on a sample from batch E256 similar to the ones described in Appendix \ref{Asec:magnetization}. The superconducting transition sets in at 4.1\,K.

\section{Selection rules for two-phonon processes and MGPT}
\label{Asec:MGPT}
In the multi-phonon scattering process the system goes from an initial vibrational state (ground vibrational state) $|0,0,\ldots \rangle$ to a final multi-phonon state $|n_{\mu},n_{\mu'},\ldots\rangle$, where $n_{\mu}$ is the number of phonons in the same state $\mu$, and $\mu$ stands for the entire set of quantum numbers (quasi-momentum $k$, angular momentum quantum number $m$, etc.). For two-phonon processes the final vibrational state is the state with two phonons in the same quantum state (double-phonon or the first overtone state), or with two phonons in different states (combination state). The corresponding matrix element for two-phonon Raman scattering is:

\begin{eqnarray}\label{Aeq:mel}
\langle 0,\ldots,n_{\mu},0,\ldots|{\cal R} |0,0,\ldots \rangle, \cr n_{\mu}=2, \textrm{overtones,}\cr
\langle 0,\ldots,n_{\mu},0,\ldots,n_{\mu'},\ldots|{\cal R} |0,0,\ldots \rangle, \cr n_{\mu}=n_{\mu'}=1, \textrm{combinations,}
\end{eqnarray}

where $\cal{R}$ is the Raman tensor. This matrix element should be a scalar, or should transform as unit representation of the system space group ${\cal S}$. The standard approximation for the Raman tensor in infinite wavelength-light approximation for the non-resonant case is the polarizability tensor, which transforms as (symmetrized) square of the vector representation, $D^{\cal{R}}({\cal S})$. Decomposition of $D^{\cal{R}}({\cal S})$ gives irreducible representations of the Raman active modes. The ground vibrational state transforms as unit representation, whereas the final two-phonon state transforms as symmetrized square, $[(D^\mu({\cal S}))^2]$, of the corresponding irreducible representation $D^\mu({\cal S})$ (overtones) or the direct product of two irreducible representations $D^\mu({\cal S}) \otimes D^{\mu'}({\cal S})$ (combinations). Symmetrization in the case of overtones comes from the bosonic nature of phonons. The matrix element (equation \ref{Aeq:mel}) transforms as reducible representation
\begin{eqnarray}
[(D^\mu({\cal S}))^2] \otimes D^{\cal{R}}({\cal S}), \textrm{ for overtones, or} \cr
D^\mu({\cal S}) \otimes D^{\mu'}({\cal S}) \otimes D^{\cal{R}}({\cal S}), \textrm{ for combinations.}
\end{eqnarray}

It is a scalar if the decomposition of the representations shown above contains the unit representation, or equivalently, if the intersection of decompositions of $[(D^\mu ({\cal S}))^2]$ or $D^\mu ({\cal S}) \otimes D^{\mu'} ({\cal S})$ and $D^{\cal{R}}({\cal S})$ is a nonempty set. To obtain selection rules for two-phonon processes, following Birman's original method,\cite{Birman1963_PR131_1489} it is enough to find the decomposition of $[(D^\mu ({\cal S}))^2]$ (for overtones) and $D^\mu ({\cal S}) \otimes D^{\mu'} ({\cal S})$ (for combinations) for all irreducible representations. If there is any representation of the Raman active mode in those decompositions then that overtone or two-phonon combination is symmetrically allowed in the Raman scattering process.
The decomposition of the (symmetrized) square of the vector representation is straightforward, and is actually a finite dimensional point group problem. On the other hand decomposition of $[(D^\mu ({\cal S}))^2]$ or $D^\mu ({\cal S}) \otimes D^{\mu'} ({\cal S})$ for any irreducible representation could be a difficult task because space groups are infinite. In the standard method based on character theory summation over all group elements is used, and it is a problem in the infinite case.
Therefore it is necessary to apply a method which avoids summation over group elements. As is proven in Ref.~\onlinecite{Damnjanovic2015_PR581_1} the modified group projector technique (MGPT) uses only group generators and finite dimensional matrices. Actually, the decomposition $D({\cal S})=\oplus_\mu f^\mu_D D^{(\mu)}({\cal S})$ of the arbitrary reducible representation $D({\cal S})$ into irreducible representations is effectively a determination of the frequency numbers $f^\mu_D$. The MGPT expression for frequency numbers involves group generators $s_i$ only:
\begin{equation}\label{Aeq:fn}
 f^\mu_{D}=
 {\mathrm {Tr}}{F\left(\prod^{S}_{i=1}F\left(D(s_i)\otimes D^{(\mu)^*}(s_i)\right)\right)}
\end{equation}
Here $S$ is the number of group generators, $F(X)$ is the projector on the subspace of the fixed points of the operator $X$, and ${\mathrm {Tr}}$ is the matrix trace (sum of the diagonal matrix elements). Consequently, the problem is reduced to calculation of the $S+1$ projector to the fixed points. Technically, one looks for the eigenspaces for the eigenvalue 1 of each of the operators $D(s_i)\otimes D^{(\mu)^*}(s_i)$, finding projectors on them, then multiplies the corresponding projectors, and repeats the procedure for the whole product from Equation~\ref{Aeq:fn}. The trace of the final projector gives the corresponding frequency number.

\begin{table*}[t]
\caption{Two-phonon processes in FeS. Symmetry group of FeS system is space group P4/nmm. For products of irreducible representations (IRs) in the left column Raman active modes (RM) in decomposition are given in the right one. Raman active modes of FeS are $\Gamma_1^+$ ($A_{1g}$), $\Gamma_2^+$ ($B_{1g}$) and two double degenerate $\Gamma_5^+$ ($E_g$). $\Gamma_1^+$ comes from vibrations of S atoms,  $\Gamma_2^+$ from Fe ones, and both atom types contribute with one pair of $\Gamma_5^+$ modes. For complex representations ($V_{1,2,3,4}$ and all $W$) double index indicates that real representation is used, for example: $V_{13}=V_1 \oplus V_1^*=V_1 \oplus V_3$. Irreducible representations of space group given in Ref. \onlinecite{Aroyo2006_ACSA62_115} are used.}
\label{table:MGPT}
\begin{ruledtabular}
\centering
\resizebox{\linewidth}{!}{%
\begin{tabular}{l c l c }
\multicolumn{2}{c}{Overtones}& \multicolumn{2}{c}{Combinations}\\ [1mm]\cline{1-2} \cline{3-4} \\[-0.5em]
\multicolumn{1}{c}{IR products} & RM  & \multicolumn{1}{c}{IR products}& RM \\
\multicolumn{1}{c}{(phonon states)}&in decomposition&\multicolumn{1}{c}{(phonon states)}& in decomposition\\ [1mm]\cline{1-2} \cline{3-4} \\[-0.5em]

$[(\Gamma_i^\pm)^2]$, ($i=1,2,3,4$)  & $A_{1 g}$ & $\Gamma_1^h\otimes \Gamma_2^h$, $\Gamma_3^h\otimes \Gamma_4^h$, ($h=\pm$)  & $B_{1 g}$ \\[1mm]
$[(\Gamma_5^\pm)^2]$  & $A_{1 g}$,$B_{1 g}$ & $\Gamma_i^h\otimes \Gamma_5^h$, ($i=1,2,3,4$, $h=\pm$)  & $E_g$ \\[1mm]
$[(X_i)^2]$, ($i=1,2$)  & $A_{1 g}$,$B_{1 g}$,$E_g$ & $X_1\otimes X_2$  & $E_g$ \\[1mm]
$[(M_i)^2]$, ($i=1,2,3,4$)  & $A_{1 g}$ & $M_1\otimes M_2$, $M_3\otimes M_4$  & $B_{1 g}$ \\[1mm]
$[(\Sigma_i)^2]$, ($i=1,2,3,4$)  & $A_{1 g}$ & $M_1\otimes M_3$, $M_1\otimes M_4$, $M_2\otimes M_3$, $M_2\otimes M_4$   & $E_g$ \\[1mm]
$[(\Delta_i)^2]$, ($i=1,2,3,4$)  & $A_{1 g}$,$B_{1 g}$ & $\Sigma_1\otimes \Sigma_2$, $\Sigma_3\otimes \Sigma_4$  & $B_{1 g}$ \\[1mm]

{$[(V_{13})^2]$, $[(V_{24})^2]$, $[(V_{5})^2]$} & { $A_{1 g}$}& $\Sigma_1\otimes \Sigma_3$, $\Sigma_1\otimes \Sigma_4$, $\Sigma_2\otimes \Sigma_3$, $\Sigma_2\otimes \Sigma_4$ & {$E_g$}\\[1mm]

$[(W_{13})^2]$, $[(W_{24})^2]$, & $A_{1 g}$,$B_{1 g}$, $E_g$ &$\Delta_1\otimes \Delta_2$, $\Delta_1\otimes \Delta_3$, $\Delta_2\otimes \Delta_4$,$\Delta_3\otimes \Delta_4$  & $E_g$\\[1mm]
$[(Y_1)^2]$  & $A_{1 g}$,$B_{1 g}$,$E_g$ & $V_{13}\otimes V_{24}$&$\Gamma_2^+$\\[1mm]
$[(Z_i^\pm)^2]$, ($i=1,2,3,4$)  & $A_{1 g}$ & $V_{13}\otimes V_5$, $V_{24}\otimes V_5$&$\Gamma_5^+$\\[1mm]
$[(Z_5^\pm)^2]$  & $A_{1 g}$,$B_{1 g}$ & $W_{13}\otimes W_{24}$&$\Gamma_5^+$\\[1mm]
$[(A_i)^2]$, ($i=1,2,3,4$)  & $A_{1 g}$ & $Z_1^h\otimes Z_2^h$, $Z_3^h\otimes Z_4^h$, ($h=\pm$)   & $B_{1 g}$ \\[1mm]
$[(R_i)^2]$, ($i=1,2$)  & $A_{1 g}$,$B_{1 g}$,$E_g$ & $Z_i^h\otimes Z_5^h$, ($i=1,2,3,4$, $h=\pm$)  & $E_g$ \\[1mm]
$[(S_i)^2]$, ($i=1,2,3,4$)  & $A_{1 g}$ & $A_1\otimes A_2$, $A_3\otimes A_4$   & $B_{1 g}$ \\[1mm]
$[(U_i)^2]$, ($i=1,2,3,4$)  & $A_{1 g}$,$B_{1 g}$ & $A_1\otimes A_3$, $A_1\otimes A_4$, $A_2\otimes A_3$, $A_2\otimes A_4$  & $E_g$ \\[1mm]
$[(\Lambda_i)^2]$, ($i=1,2,3,4$)  & $A_{1 g}$ & $R_1\otimes R_2$  & $E_g$ \\[1mm]
$[(\Lambda_5)^2]$  & $A_{1 g}$,$B_{1 g}$ & $S_1\otimes S_2$,$S_3\otimes S_4$  & $B_{1 g}$ \\[1mm]
$[(T_1)^2]$  & $A_{1 g}$,$B_{1 g}$,$E_g$ & $S_1\otimes S_3$, $S_1\otimes S_4$, $S_2\otimes S_3$, $S_2\otimes S_4$  & $E_g$ \\[1mm]
&& $U_1\otimes U_2$, $U_1\otimes U_3$, $U_2\otimes U_4$, $U_3\otimes U_4$  & $E_g$ \\[1mm]
&& $\Lambda_1\otimes \Lambda_2$, $\Lambda_3\otimes \Lambda_4$  & $B_{1 g}$ \\[1mm]
&& $\Lambda_i\otimes \Lambda_5$, ($i=1,2,3,4$)  & $E_g$ \\[1mm]

\end{tabular}}
\end{ruledtabular}
\end{table*}

\end{appendix}

\end{document}